\documentclass[aps, twocolumn, nofootinbib, showpacs, prd]{revtex4}
\usepackage{graphicx}
\usepackage{amsfonts}
\usepackage{amssymb}
\usepackage{amsbsy}
\usepackage{amsmath}
\usepackage{latexsym}
\usepackage{natbib}
\usepackage{bm}
\usepackage{color}
\usepackage{float}

\begin{document}
\title{Scalar-Fermion Interaction as the Driver of Cosmic Acceleration}

\author{Soumya Chakrabarti\footnote{soumya1989@bose.res.in}}
\affiliation{Department of Theoretical Sciences,\\
S. N. Bose National Centre for Basic Sciences, \\
JD Block, Sector-III, Salt Lake City, Kolkata - 700 106, India
}

\author{Amitabha Lahiri\footnote{amitabha@bose.res.in}}
\affiliation{Department of Theoretical Sciences,\\
S. N. Bose National Centre for Basic Sciences, \\
JD Block, Sector-III, Salt Lake City, Kolkata - 700 106, India
}
 
\pacs{}

\date{\today}

\begin{abstract}
We argue that an interacting scalar-fermion distribution can be used to demonstrate the cosmic acceleration in General Relativity. The interaction is of Yukawa nature and it drives the fermion density to decay with cosmic time. The consistency of the model is established through, (a) a generalization of the $Om(z)$ parameter, the present matter density contrast of the universe and (b) a comparison of theoretical results with observational data, using a Markov Chain Monte Carlo code. A simple model of unified cosmic expansion history is also discussed. There are more than one instance where the universe goes through a transition between subsequent epochs with different patterns of cosmic expansion. These patterns impose some constraints on the scalar-fermion interaction profile and on the overall cosmological dynamics.
\end{abstract}

\maketitle

\section{Introduction}
The field equations of General Relativity (GR) relate the curvature of space-time with the sources of energy-momentum distributions in the universe. Most of the known gravitational phenomena are determined by exact or numerical solution of these equations. However, quite a few outstanding riddles remain in this context. One of them is the lack of explanation to the accelerated expansion of our universe. This acceleration is indeed confirmed by a range of astrophysical observations~\cite{perlmutter, riess0, melchiorri, jaffe, lange, halverson} and therefore, should inspire a credible theoretical explanation, even if the standard gravitational force is an attractive force. The observations also demand that there be an era of deceleration immediately before the current epoch of acceleration, such that the expansion slows down just enough to allow a consistent formation of galaxies~\cite{riess1, paddy1}. While standard GR can not, on its own support this, speculation is that one needs extended theories for an observational consistency. These extensions are often motivated by theories of fundamental interactions beyond GR and they introduce a few \textit{exotic} constituents in our universe, for example, the \textit{cold dark matter} particles to stabilize galactic structures, and the \textit{dark energy} fluid to produce the desired late-time acceleration. A consistent transition between subsequent epochs can be ensured using kinematic quantities such as the deceleration parameter~\cite{paddy1, paddy2}. There exist quite a few extensions of standard cosmology, either in the curvature sector or in the energy-momentum source. They produce an evolving effective Equation of State (EOS) that can evolve into the present era as a generator of a negative pressure. All such efforts have enriched the subject of gravitation and cosmology for some decades now, and in a combined manner, they spearhead the search for the Dark Sectors of the Universe~\cite{copeland}.  \\

The simplest entity that can fill in for a Dark Energy fluid is the Cosmological Constant~\cite{riess2, eisenstein}, although there is no fully acceptable mechanism for generating such a small constant ($\sim 10^{-120} l^{-2}_{P}$). Quintessence, a particular kind of self-interacting scalar field can be a potential resolution, as it can produce a time evolving energy density correction~\cite{zlatev, sahni}. However, the so-called \textit{fifth force experiments} impose tight constraints on the nature of interaction between a Quintessence scalar and the baryon content of the universe, ultimately dismissing their viability~\cite{adelberger}. Some specific cases of interest do exist such as the pseudo-Nambu-Goldstone-Boson~\cite{frieman}. There is another possibility that the scalar field driving cosmic acceleration avoids being detected in the Fifth Force experiments, by decoupling in regions of higher density due to a `screening mechanism' (e.g. a chameleon)~\cite{khoury, hinterbichler}. This is not a straightforward construct as the screening mechanisms can face a problem due to the mass scale of the scalar field~\cite{wang-hui-khoury}. A better resolution comes from scalar-tensor theories of gravity which include geometric scalar fields. Brans-Dicke theory is the archetype of these theories, which allows a variation of gravitational coupling~\cite{bransdicke}. Although the standard Brans-Dicke theory suffers from strict bounds from local astronomical tests and falls short of providing a viable cosmological alternative, generalizations of the theory can still provide good phenomenological explanations~\cite{guth, mathiazhagan, la, banerjee1}. There is always a possibility that further extensions of the theory can turn out to be a rejuvenation~\cite{nordtvedt, banerjee2, banerjee3, faraoni1, bertolami1, sen}. Recent works in this regard show that an interacting Higgs scalar field in a generalized Brans-Dicke theory can act as the driver of cosmological expansion \cite{chakrabarti1} and at the same time explain the cosmic variation of Higgs vacuum expectation value \cite{sola, chakrabarti2}.  \\

The greater problem in cosmology, however, is to identify \textit{the best modification} amongst a plethora of examples. This modification should describe a unified time history, desirably starting with an early inflation followed by subsequent epochs of decelerated expansion and late-time acceleration. Well-known dark energy candidates, such as exotic fluids with non-trivial equations of state (EOS) or scalar fields with different self-interaction profiles have received their due attention in this issue (see for instance the review~\cite{copeland}). However, they can only explain the phenomenology partially and it is desirable to keep exploring different angles for a better description. This is where a fermionic field can come in as an interesting alternative~\cite{obukov, saha1, saha2, saha3, saha4, ribas3, desouza, samojeden, saha5, cai1, cai2, wang}. There are examples where Fermionic matter is considered as a candidate, separately, for early acceleration as well as the dark energy. The so-called `old universe' models, showing a transition into fermion dominated acceleration from a matter dominated deceleration, deserves mention in this context. However, bringing in an ever-present cosmologically responsible Fermionic field is a little radical, since by definition it can exhibit negative Casimir or vacuum energy (direct consequence of Lorentz invariance and Fermi-Dirac statistics). The idea of an interacting mix of Bosonic and Fermionic fields driving the cosmic acceleration can work as an interesting yet diplomatic approach to the issue, provided it produces an evolving EOS and satisfy constraints from astrophysical observations~\cite{ribas1, ribas2, chimento}. This approach has also received some attention in the context of early universe cosmology, for example, to produce inflationary attractor solutions~\cite{domenech}, and as a probable generator of primordial black holes~\cite{domenech}. Interacting boson-fermion models can also provide an alternative insight into the dark energy-dark matter interaction in late-time universe and in turn, produce an outline of structure formation~\cite{amendola, farrar}. This mechanism also motivates the interacting quintessence-scalar-neutrino setup, which can explain the time variation of neutrino masses~\cite{amendola, wetterich, schech}. There is, however, an issue of incompleteness of these models, owing to their non-relativistic framework~\cite{bentoconcl, casas, mohseni}. A relativistic formalism of the boson-fermion interaction, even at the level of a toy model, is immensely important, and will be considered at some length in this manuscript. Further generalizations should involve dilatonic interactions~\cite{amendola1} (similar to geometric couplings in string theory or GR~\cite{fujii, jordan, bd}) and will be addressed in the near future in a separate work. \\

In this work, we consider a Yukawa-type interaction between a fermionic field and a scalar field. The fermion has a quartic self-interaction while the scalar self-interaction profile can be chosen at will. A similar model has received some attention in literature~\cite{ribas1, ribas2, ribas3}, however, the question of cosmological consistency based on credible observational data remains unaddressed as yet. We discuss the structure of the model such that standard phenomenological requirements of a smooth transition from deceleration into the recent phase of acceleration can be met. This is done through a cosmological reconstruction from the $Om(z)$ parameter~\cite{sahni1, sahni2, lu, tong}. Broadly, there are two popular methods to perform a cosmological reconstruction. A parametric reconstruction involves a pre-assumed parametric form of one or more cosmological quantities and comparison with a set of observational data~\cite{chen, ryan}. On the other hand, a non-parametric reconstruction involves direct statistical analysis of sets of observational data, for instance, through a principal component analysis~\cite{crittenden, clarkson, amendolaetal, holsclaw, shafieloo}. Parametric reconstruction techniques based on kinematic quantities have become much popular lately owing to their simplicity \cite{bernstein, visser, cattoen, dunajski}. The most popular choices for this purpose are the deceleration and jerk parameters \cite{ankan}, involving second and third order derivatives of the scale factor respectively. However, higher order kinematic quantities such as the statefinder parameter~\cite{alam11}, also seem to work really well. They can help one compare and determine the departure of an envisaged Dark Energy model from the standard $\Lambda$CDM cosmology. $Om(z)$ primarily became popular as the centerpiece of many successful attempts at observational data analysis and subsequent comparisons of modified dark energy models~\cite{sahni1, sahni2}. It is essentially a constant defining the present matter density of the universe. We parametrize this definition to find an observationally viable Hubble as a function of redshift. Thereafter, we use it to solve the field equations of the theory and discuss the best possible structure. We also discuss a toy model, describing how this interacting boson-fermion mix might behave in a unified evolution of the universe. This is done by using a solution of the cosmological scale factor as a function of cosmic time, which successfully describes an early acceleration, followed by an epoch of extended deceleration and finally, the late-time acceleration. \\

We have used the metric signature $(+,-,-,-)$ and natural units with $8\pi G=c=\hbar=1$. In the next Section, the basic mathematical formalism of an interacting boson-fermion fluid is dicussed. We outline the  scheme of reconstruction in Section $3$. The best possible structure of the theory is discussed in Section $4$. We include the discussion on a unified cosmological evolution scenario in Section $5$, and conclude in Section $6$. 

\section{Mathematical Formulation}
In order to study fermions in curved spacetime, we need to use the tetrad formalism. A tetrad $e^a_\mu$ is a section of the frame bundle, i.e., a set of four orthonormal basis vectors at each point. It relates the spacetime metric $g_{\mu\nu}$ with the \textit{internal} Minkowski metric $\eta_{ab}$ by 
\begin{equation}
g_{\mu\gamma} = e^a_\mu e^b_\gamma\eta_{ab}\,.
\end{equation}
We can think of the tetrad as a linear isomorphism between the internal flat space and the tangent space at each point. The co-tetrad $e^\mu_a$ is the inverse tetrad defined by the relation $e^\mu_a e^a_\nu = \delta^\mu_\nu\,,$ such that 
\begin{equation}
\eta_{ab} = g_{\mu\nu}e^\mu_a e^\nu_b\,.
\end{equation}
Greek indices refer to the spacetime and the Latin indices refer to the internal space. $a = 0, 1, 2, 3$. The Dirac matrices $\gamma^a$ are defined on the internal flat space, where we define the local matrices as
\begin{equation}
\Gamma^{\mu} = e^\mu_a\gamma^a\,,
\end{equation}
satisfying the anticommutation relation
\begin{equation}
\{\Gamma^\mu,\Gamma^\nu\} = 2g^{\mu\nu}\,.
\end{equation}
The covariant derivative of a Dirac field is written as 
\begin{equation}
D_\mu\psi= \partial_\mu\psi-\Omega_\mu\psi\,,\qquad
D_\mu\overline\psi=\partial_\mu\overline\psi+\overline\psi\Omega_\mu\,,
\label{covder}
\end{equation}
where $\bar\psi = \psi^\dagger\gamma^0\,$. The spin connection $\Omega_\mu$ corresponds to torsion. The gravitational action then also contains torsion, but only as $T^2$ after neglecting a total divergence. The equation of motion sets the torsion equal to a fermion current which can be put back into the action in a form of a four-fermion interaction term~\cite{Hehl:1971qi, Gasperini:2013, Chakrabarty:2018ybk}. The fermion action is then written in terms of the torsion-free spin connection
\begin{equation}
\omega_\mu=-\frac{1}{4}g_{\rho\sigma}[\Gamma^\rho_{\mu\delta}-e_b^\rho(\partial_\mu e_\delta^b)]\Gamma^\delta\Gamma^\sigma\,.
\label{spinconn}
\end{equation}
$\Gamma^\nu_{\sigma\lambda}$-s are the usual Christoffel symbols. 
The covariant derivative of a bosonic field, as well as the connection in the action, are torsion-free. In general, the four-fermion interaction contains products of vector and pseudo-vector currents~\cite{Chakrabarty:2019cau, Lahiri:2020mst}. This can be rewritten for a single species of fermion as products of scalar and pseudo-scalar bilinears by use of Fierz identities. For the purpose of this toy model, we include a general self-interaction term $V(\psi)\,$ and write the Lagrangian density of a Dirac field in curved spacetime as 
\begin{equation}
\mathcal{L}_{\rm Dirac}(\psi)=\frac{\imath}{2}[ \overline\psi\,\Gamma^\mu D_\mu\psi-(D_\mu\overline\psi)\Gamma^\mu\psi]-V(\psi).
\label{Dirac-L}
\end{equation}
We note that in a more realistic model, there will be a sum over different species of fermions, and $V(\psi)$ will include current-current and other interaction terms between different species. In this paper we will keep things simple and work only with one species of fermion. The scalar field is interacting with the fermion through a Yukawa interaction. The scalar field is also self-interacting, although we do not specify the interaction at this point. The scalar field contribution to the Lagrangian is 
\begin{equation}
\mathcal{L}_{\rm scalar}(\phi)={1\over2}\partial^\mu\phi\,\partial_\mu\phi - U(\phi) - \lambda\overline\psi\phi\psi,
\end{equation}
where $\mathcal{L}_{\rm Yukawa}(\phi,\psi) = -\lambda\overline\psi\phi\psi$ is the Yukawa interaction. Denoting the Ricci curvature by $R$ as usual, we write the action for this scalar-fermion system in natural units as
\begin{align}
S =&\int d^4x\sqrt{-g}\left\{\frac{1}{2}R + {1\over2}\partial^\mu\phi\,\partial_\mu\phi - U(\phi) -\lambda\overline\psi\phi\psi\right. \notag
\\\label{action}
&\left.+\frac{\imath}{2}\left[\overline\psi\Gamma^\mu D_\mu\psi-(D_\mu\overline\psi)\Gamma^\mu\psi\right]-V(\psi) + L_{m} \right\}\,.\qquad
\end{align}
Note that there is an additional fluid distribution $L_m$ in the action. The fluid is non interacting and is expected to contribute to the cold dark matter distribution in some manner. Moreover, this provides additional dynamical variables to the system through the components of fluid energy momentum tensor. 

The fermion field $\psi$ and its adjoint $\bar\psi$ obey the equations
\begin{eqnarray}\label{psieq}
&& 
\imath\Gamma^\mu D_\mu\psi-{\partial V\over \partial{\overline\psi}}-\lambda\phi\psi= 0\,, 
\\
&& \label{barpsieq} \imath D_\mu\overline\psi\,\Gamma^\mu+{\partial V\over \partial\psi}+\lambda\overline\psi\phi= 0\,.
\end{eqnarray}
Variation of the action with respect to the scalar field $\phi$ leads to
\begin{equation}\label{K-G}
\nabla_\mu\nabla^\mu\phi + \frac{dU}{d\phi} + \lambda\overline\psi\psi=0\,,
\end{equation}
while varying with respect to the metric leads to the Einstein equations
\begin{equation}\label{EE}
R_{\mu\nu}-\frac{1}{2}g_{\mu\nu}R = -T^{\rm eff}_{\mu\nu}.
\end{equation}
$T^{\rm eff}_{\mu\nu}$ is the effective energy-momentum tensor. We can calculate the scalar and fermion contributions to the energy-momentum tensor from the metric variation of the terms involving $\psi$ and $\phi$
\begin{eqnarray}\nonumber
T^{\mu\nu} &=& \frac{\imath}{4}\left[\overline\psi\Gamma^{(\mu} D^{\nu)}\psi - D^{(\mu}\overline\psi\Gamma^{\nu)}\psi\right] + \partial^\mu\phi\partial^\nu\phi \\
&&\, -g^{\mu\nu}\left[{1\over2}\partial^\sigma\phi\,\partial_\sigma\phi-U(\phi) 
-\lambda\overline\psi\phi\psi \right. \nonumber \\
&&\left.
+\frac{\imath}{2}\left(\overline\psi\Gamma^\lambda D_\lambda\psi-D_\lambda\overline\psi\Gamma^\lambda\psi\right)-V(\psi)\right]\,.
\end{eqnarray}

We now consider this system on the background of a spatially homogeneous and isotropic Friedmann-Lema\^itre-Robertson-Walker universe with the line element 
\begin{equation}
ds^2 = dt^2-a^2(t)(dx^2+dy^2+dz^2)\,, \label{11}
\end{equation}
$a(t)$ being the scale factor which is a function of cosmic time. The tetrad components, the spin connection and the Dirac matrices can be simplified for this geometry as
\begin{eqnarray}
&& e_0^\mu=\delta_0^\mu\,,\qquad e_i^\mu=\frac{1}{a(t)}\delta_i^\mu\,, \\ 
&& \Gamma^0=\gamma^0\,, \qquad 
\Gamma^i=\frac{1}{a(t)}\gamma^i\,, \\
&& \Omega_0=0\,,\qquad  \Omega_i=\frac{1}{2}\dot a(t)\gamma^i\gamma^0\,.
\end{eqnarray}
In this geometry the scalar and the fermionic field evolve only with time. Then Eqs. (\ref{psieq}), (\ref{barpsieq}) and (\ref{K-G}) simplify into
\begin{eqnarray}\label{8a}
&& \dot\psi+\frac{3}{2}\frac{\dot a}{a}\psi+\imath\gamma^0{\partial V\over \partial{\overline\psi}}+\imath \lambda\gamma^0\psi\phi=0,\\&& \label{8b}
\dot{\overline\psi}+\frac{3}{2}\frac{\dot a}{a}\overline\psi-\imath{\partial V\over \partial\psi}\gamma^0-\imath\lambda\phi\overline\psi\gamma^0=0,\\&& \label{8c}
\ddot\phi+3\frac{\dot a}{a}\dot\phi+ \frac{dU}{d\phi} +\lambda\overline\psi\psi=0.
\end{eqnarray}

We multiply Eq. (\ref{8a}) with $\overline\psi$ on the left and Eq. (\ref{8b}) with $\psi$ on the right to eliminate the $\imath \lambda{\overline\psi}\gamma^0\psi\phi$ term. This produces a simple differential equation for the bilinear $\overline\psi\psi$ as follows
\begin{equation}\label{steps}
\frac{d}{dt}({\overline\psi}\psi) + 3\frac{\dot{a}}{a}({\overline\psi}\psi) + \imath  \left(\overline\psi\gamma^0{\partial V\over \partial{\overline\psi}} - {\partial V\over \partial{\psi}}\gamma^0\psi \right) = 0.
\end{equation}
Note that for specific choices of the fermionic self-interaction we can integrate the above equation and write the bilinear $\overline{\psi}\psi$ as a function of the cosmological scale factor. We focus on potentials of the form
\begin{equation}
V = \delta_{0}\left(\overline\psi\psi\right)^n\,,
\end{equation}
where $\delta_{0}$ is a coupling constant. The parameter $n$ signifies the strength of the self-interaction. With this assumption Eq. (\ref{steps}) can be integrated to give
\begin{equation}
\overline\psi\psi=\frac{C}{a^3}\,,
\end{equation}
where $C$ is a constant of integration. An interesting point in this regard is that we can also derive this exact form of the bilinear for a more generalized fermionic potential such as
\begin{equation}
V = \sum_{i=0}^{n} \delta_{i}\left(\overline\psi\psi\right)^i \,.
\end{equation}

Replacing the bilinear as a function of scale factor, we write the independent cosmological equations as
\begin{align}
 \ddot{\phi} &+ 3H\dot{\phi} + \frac{dU}{d\phi} + \frac{C\lambda}{a^3} = 0\,, \label{kg1}
 \\
3H^{2} &= \rho_{m} + \frac{\dot{\phi}^2}{2} + U(\phi) + \frac{C\lambda \phi}{a^3} + \frac{C \delta_{0}}{a^{3n}}\,, \label{kg2} \\ 
-2\dot{H} - 3H^{2} & = p_{m} + \frac{\dot{\phi}^2}{2} - U(\phi) + \frac{(n-1)C \delta_{0}}{a^{3n}}\,. \quad \label{kg3}
\end{align}

These equations define the system. If we assume a form of the scalar self-interaction, we technically have three independent equations and four unknown quantities to solve for, namely, the Hubble function, the scalar field, the density and the pressure of the fluid. While more common approaches are to take some form of the scalar field or a pre-assigned EOS of the accompanying fluid, we chose to solve these equations with a reconstructed form of Hubble. The reconstruction scheme is discussed in the next section.

\section{A Reconstruction from {$Om(z)$} Parameter}
A cosmological reconstruction is a method of finding a `best possible structure' of any theory, based on a set of desired cosmological behavior. This practice often promotes the use of purely kinematic quantities, which are written as combinations of the scale factor and its higher order derivatives. In this work, we apply a simple scheme of reconstruction using the $Om(z)$ parameter~\cite{alam11}, which is written as a simple combination of the redshift and the Hubble function. It has recently gained popularity in the form of an \textit{`Om diagnostic'}. Combined with a statefinder diagnostic it contributes in large scale cosmological data analysis, both in a model dependent and model dependent manner. A parametrization of $Om(z)$ can put reasonable constraints on the estimate of present matter density contrast in a modified cosmological scenario. For a standard $\Lambda$CDM model, $Om(z)$ is a constant, written as $\Omega_{m0}$. Generalizing this, we write the parameter as a function of redshift. The standard expression for the $Om(z)$ parameter is
\begin{align}
Om(z) &= \frac{h(z)^{2}-1}{(1+z)^{3}-1}\,, \\
{\rm where}\qquad h(z) &= \frac{H(z)}{H_{0}}\,.
\end{align}
$H_{0}$ is the present value of the Hubble parameter. A generalized version, $Om(z)$, is written as
\begin{equation}\label{ansatz}
Om(z) = \Omega = \lambda_{0} (1 + z)^{\delta}.
\end{equation}
This leads to
\begin{equation}\label{hubbleansatz}
h(z) = \left[1 + \lambda_{0} (1+z)^{\delta} \left\lbrace (1+z)^{3} - 1 \right\rbrace\right]^{\frac{1}{2}}.
\end{equation}
$h(z) = H(z)/H_{0}$ is the dimension-less Hubble, including a scaling of $H_0$ by $100$ km $\mbox{Mpc}^{-1}$ $\mbox{sec}^{-1}$, which is more suitable for a comparison with observational data. 

\begin{table*}[t!]
\caption{{\small Best Fit Parameter values of three parameters : (i) Dimensionless Hubble $h_{0}$, (ii) departure from $\Lambda$CDM $\delta$ and (iii) present matter density contrast $\lambda_0$. Estimations of 1$\sigma$ uncertainty are also given.}}\label{resulttable}
\begin{center}
{\centering
\begin{tabular}{l  c  c  c  c  c  c  c}
\hline 
 & \multicolumn{1}{c}{$h_0$} & \multicolumn{1}{c}{$\delta$} & \multicolumn{1}{c}{$\lambda_{0}$} \\
\hline \\ 
$OHD+JLA+BAO$ 	  & $0.704^{+0.007}_{-0.007}$ &$0.052^{+0.038}_{-0.037}$ & $0.256^{+0.020}_{-0.019}$ & \\ \, \\
\hline
\end{tabular}
}
\end{center}
\end{table*}

\begin{figure}[t!]
\begin{center}
\includegraphics[angle=0, width=0.52\textwidth]{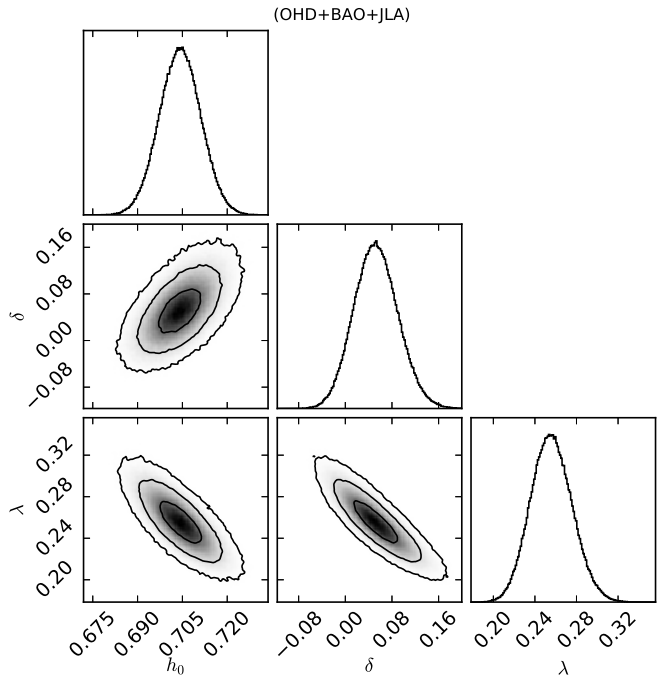}
\caption{Parameter Space Confidence Contours showing estimation of the uncertainty, the best fit and the likelihood analysis of parameters (combination of data from OHD+JLA+BAO). Associated 1$\sigma$ and 2$\sigma$ confidence contours are also shown.}
\label{Modelcontour}
\end{center}
\end{figure}

The very basic requirements of an observationally consistent cosmological dynamics are (i) a present value of the Hubble and deceleration parameter close to direct observations, (ii) an evolution of Hubble function and the deceleration that satisfies the Luminosity distance measurement data and a transition of the universe without any discontinutiy. The model parameters must be estimated such that these requirements are met and we use a set of observations - the Supernova distance modulus data (Joint Light Curve Analysis from $SDSS-II$ and $SNLS$ collaborations)~\cite{betoule}, the direct measurement of present value of Hubble parameter (OHD) \cite{simon, stern, blake, moresco, chuang, planck, delubac} and the Baryon Acoustic Oscillation (BAO) data from the BOSS collaborations and $6dF$ $Galaxy$ $Survey$ \cite{beutler, boss}. For the estimation we use a statistical analysis and present the results in the form of confidence contours on the parameter space. The contours point out to the best fit values of of three parameters : (i) Dimensionless Hubble $h_{0}$, (ii) departure from $\Lambda$CDM, signified by $\delta$ and (iii) the present matter density contrast $\lambda_0$. The numerical code used is a \textit{Markov Chain Monte Carlo simulation} (MCMC) \cite{foreman} written in python. We present the best possible values of the three parameters and 1$\sigma$ error estimations in Table.~\ref{resulttable} for a clear reference. \\

\begin{figure}[t!]
\begin{center}
\includegraphics[angle=0, width=0.40\textwidth]{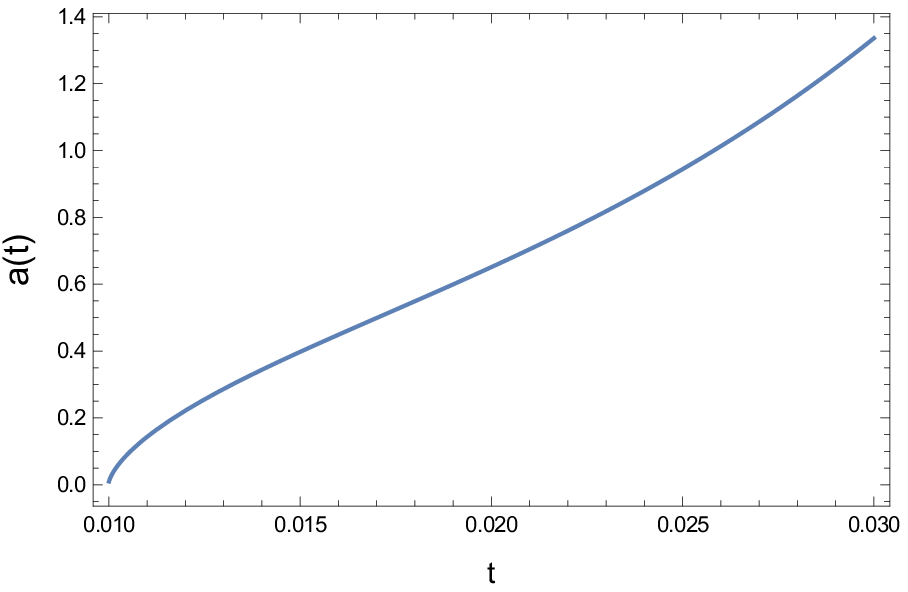}
\includegraphics[angle=0, width=0.40\textwidth]{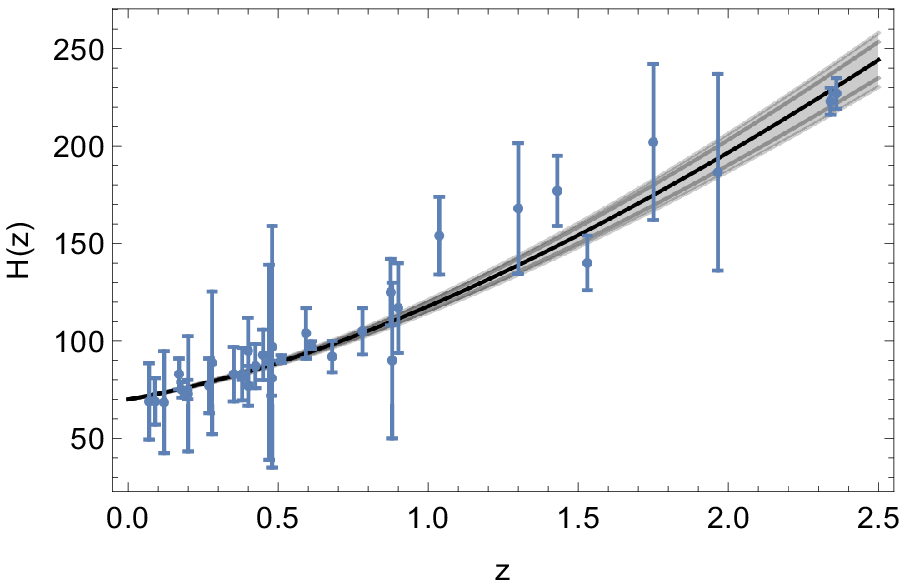}
\caption{Evolution of the reconstructed Hubble Function. The thick black line is for best fit parameter values and the gray regions are for associated 2$\sigma$ and 3$\sigma$ confidence regions. The data points of Supernova distance modulus are fitted with the curve}
\label{Hz_data}
\end{center}
\end{figure}

$H_{0}$, the current value of Hubble parameter is quite consistent with the observations of Planck \cite{planck}. There is, however, a clear departure from a standard $\Lambda$CDM model, for which the parameter $\delta$ is expected to be $0$. The present cosmological dynamics, due to the parametrization leads to a model with $\delta$ in the range of $0.052^{+0.038}_{-0.037}$. This departure can be used as a signature while studying the nature of Dark Energy and identify it as Quintessence-like or Phantom-like \cite{shafieloo1, wang1}. However, this seems a small departure at this moment and even if it does not contribute significantly in standard cosmological evolutions, it should not be ignored. The evolution of $H(z)$ as a function of $z$ is shown in Fig.~\ref{Hz_data}. The data points of Supernova distance modulus are fitted with the curve and the ranges of associated uncertainty. It shows a sufficient match during the late-time epoch. We also plot the scale factor as a function of cosmic time, within an arbitrary reference of time-scale, showing an early deceleration followed by the late time acceleration.

\begin{figure}
\begin{center}
\includegraphics[angle=0, width=0.40\textwidth]{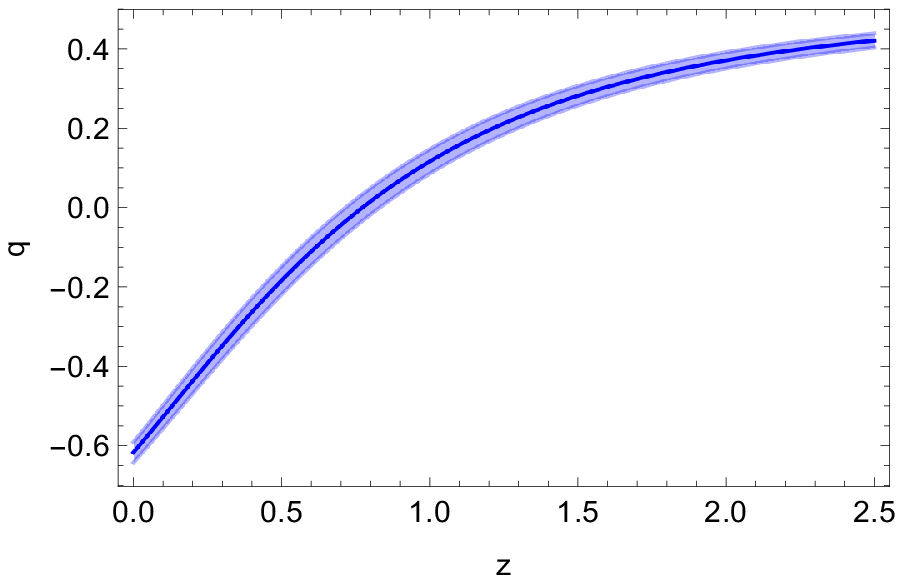}
\includegraphics[angle=0, width=0.40\textwidth]{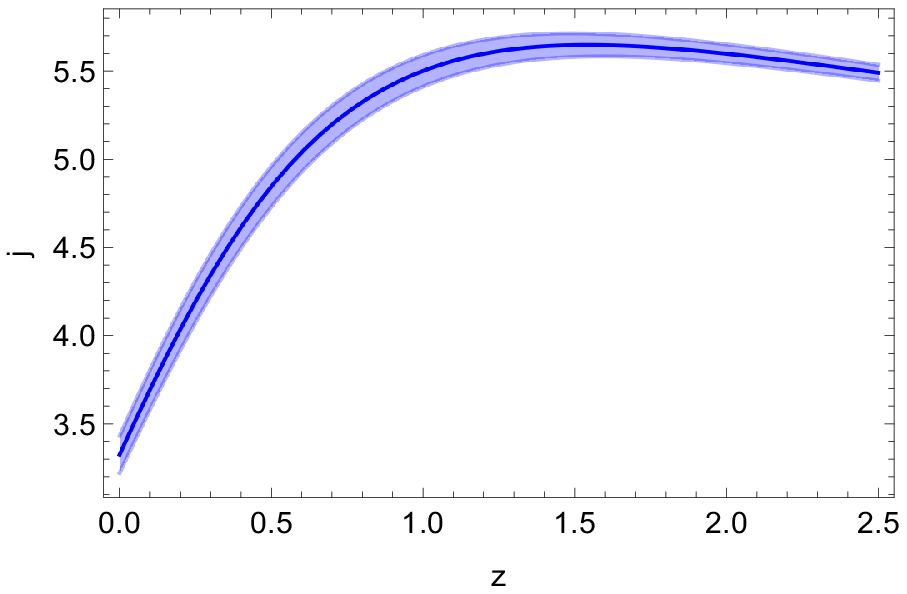}
\caption{Evolution of the deceleration parameter as a function of redshift on the top and the jerk parameter as a function of redshift on the bottom. The bold curve is for best fit values and the faded regions are for associated confidence estimation region.}
\label{cosmic_parameters}
\end{center}
\end{figure}

Fig.~\ref{cosmic_parameters} shows the plots of deceleration and jerk parameters as functions of the redshift. These evolutions are for the choice of best fit parameter values of $H_{0}$, $\lambda_{0}$ and $\delta$. The mathematical definition of these parameters are given below
\begin{align}\label{eq:eq1.2}
&q = -\frac{\ddot{a}a}{\dot{a}^2} = -\frac{\dot H}{H^2}-1, \\ & 
j = \frac{\ddot H}{H^3}+3\frac{\dot H}{H^2}+1. \label{eq:eq1.3}
\end{align}
The bold blue plot shows the evolution for best fit values and the faded blue regions are for associated regions of uncertainty. We estimate that the present value of deceleration $q(z)$ is $\sim -0.62$, which affirms the fact that the present universe is expanding with a negative deceleration or an acceleration. The present value is very well consistent with observations. From Fig.~\ref{cosmic_parameters} we see that a smooth transition from deceleration into acceleration happens around some redshift $z_{t} < 1$. This too, goes very well with observational expectations. We plot the evolution of jerk parameter in the bottom graph of Fig.~\ref{cosmic_parameters} and see that it does exhibit a departure from $\Lambda$CDM (for which jerk is always $1$~\cite{copeland}). \\ 

It is crucial that the late time evolution of a reconstructed universe should allow a consistent growth of matter over-density. Due to gravity there is a cosmological growth of matter over-density through mass accumulation from the surrounding. This growth is related to the background matter density and the universe expansion rate. On a large scale, structure formation is well described by a $\Lambda$CDM growth and therefore, a modified description of cosmology should also exhibit a behavior close to this. The definition of matter density contrast is
\begin{equation}
\delta_m = \frac{\delta\rho_m}{\rho_m}.
\end{equation}
We have assumed the background matter density $\rho_m$ to be homogeneous while any deviation is written as $\delta\rho_m$. $\delta\rho_m$ evolves non-linearly in a region of strong gravity, for instance, the over-dense region around a collapsing stellar distribution. However, for a spatially homogeoneous cosmological system the non-linear evolution can be approximated by a linearized equation~\cite{linder, percival, pace} 
\begin{equation}\label{delm_eq}
\ddot{\delta}_m + 2H\dot{\delta}_m = 4\pi G\rho_m\delta_m.
\end{equation}

\begin{figure}
\begin{center}
\includegraphics[angle=0, width=0.40\textwidth]{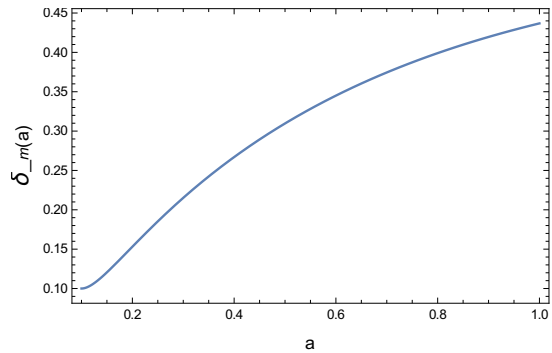}
\caption{Growth pattern of matter overdensity $\delta_m$ with scale factor $a$ for the best fit parameter values.}
\label{overdensity}
\end{center}
\end{figure}

We rewrite this equation using the scale factor as an independent variable instead of cosmic time and solve it numerically for $\delta_m$. The initial conditions for scale factor, over-density and its first derivative are taken to be $a_i = 0.01$, $\delta_m(a_i) = 0.01$ and $\dot{\delta}_m(a_i) = 0$. The numerical solution is given as a plot of $\delta_m$ vs $a$ in Fig.~\ref{overdensity}, for the best fit parameter values. The reconstructed model appears to follow a growth pattern of matter over-density that is closely follows a $\Lambda$CDM behavior (see \cite{linder} for more discussions). Therefore, the departure from standard cosmology, in its present scale, do not affect the matter overdensity growth pattern in a significant manner. \\

We also discuss the thermodynamic equilibrium of this reconstructed cosmological system. The idea is easier to conceive if one imagines the expanding universe to be a thermodynamic system within a boundary and compares it with a black hole in thermodynamic equilibrium~\cite{GibbonHawking, Jacobson:1995ab}. We call the boundary to be a cosmological horizon, which, for a spatially flat cosmology is known as the Hubble horizon~\cite{Bak:1999hd}, defined as,  
\begin{equation}
r_h = (H^2+k/a^2)^{-1/2}.
\end{equation}
Thus for $k=0$, $r_h = 1/H$. Defining this horizon is crucial for the laws of black hole thermodynamics to hold. The total entropy of the horizon encircled system $S$ is a sum of entropy contributions from the boundary and the constituent fluid elements,
\begin{equation}
S=S_f+S_h.
\end{equation}
There must be two constraints on the total entropy throughout the cosmic expansion history, (i) it should not decrease with cosmic time and (ii) it can not have a minima~\cite{GibbonHawking}. Mathematically,
\begin{align}\label{thermoreq.1}
 \frac{dS}{dn} &\geq 0, \\
\frac{d^2S}{dn^2} &< 0. \label{thermoreq.2}\\
\nonumber
n&=\ln{a}.
\end{align}

The entropy contribution due to the horizon is related to the horizon area ${\mathcal A}=4\pi r^2_h$ and in a unit where $\hbar = k_B = c = 8\pi G = 1$ is written simply as
\begin{equation}
S_h=8\pi^2 r^2_h.
\end{equation}
We assume that constituent fluid elements within the volume $V = 4\pi r^3_h/3$ have a uniform temperature $T$. Moreover, it is also argued that the horizon contribution to the temperature, $T_h$, is inversely proportional to the radius $r_h$~\cite{Jacobson:1995ab, Bak:1999hd, Frolov:2002va}. Assuming the dominant cosmic fluid components to be dark energy (de) and the pressureless cold dark matter (cdm), we write the effective fluid entropy as
\begin{equation}
S_f = S_{cdm} + S_{de}.
\end{equation}

This leads to the first law of thermodynamics written as
\begin{align}
 TdS_{cdm}&=dE_{cdm}+p_{m}dV=dE_{cdm}, \\
TdS_{de}&=dE_{de}+p_{d}dV.
\end{align}
The energy contributions of these two components within the sphere of radius $r_h$ can be written as
\begin{align}
E_{cdm} &=\frac{4\pi r^3_h\rho_{cdm}}{3}, \\
E_{de} &=\frac{4\pi r^3_h\rho_{de}}{3}.
\end{align}

We further assume that the temperature is uniform through fluid and horizon, i.e., $T = T_{h}$. Then the first time derivative of entropy can be written as
\begin{equation}
\dot{S}=\dot{S}_{cdm}+\dot{S}_{de}+\dot{S}_{h}=4\pi Hr_h^2[\rho_m+(1+w_{de})\rho_{de}]^2.
\label{Sdot}
\end{equation}
To consider the condition for a thermodynamic equilibrium, it is better to write Eq. (\ref{Sdot}) as a function of scale factor, or $n = \ln{a}$. After a few straightforward algebraic mainipulations, the first and second order change can be written as a function of Hubble
\begin{align}\notag
S_{,n} &= \frac{16\pi^2}{H^4}(H_{,n})^2, \\ \label{Sdn}
S_{,nn} &= 2S_{,n}\left(\frac{H_{,nn}}{H_{,n}}-\frac{2H_{,n}}{H}\right).
\end{align}

\begin{figure}
\begin{center}
\includegraphics[width=0.40\textwidth]{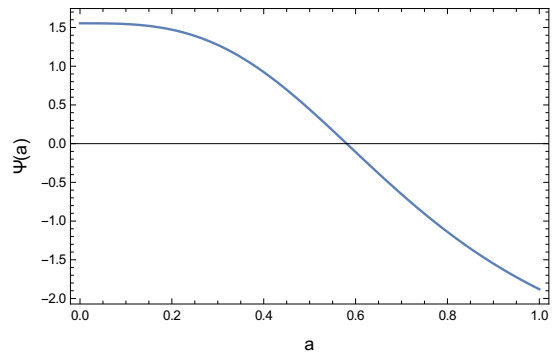}
\caption{Pattern of evolution of the $\Psi$-factor with scale factor, $\Psi = \left(\frac{H_{,nn}}{H_{,n}}-\frac{2H_{,n}}{H}\right)$.}
\label{psiplot}
\end{center}
\end{figure}

For convenience, we write $\left(\frac{H_{,nn}}{H_{,n}} - \frac{2H_{,n}}{H}\right) = \Psi$. It is now easy to note from Eq. (\ref{Sdn}) and Eq. (\ref{thermoreq.2}) that a thermodynamic equilibrium requires the $\Psi$-factor to be negative. We can easily check this for the reconstructed model using the closed form of $H(z)$ as in Eq. (\ref{hubbleansatz}). A $\Psi$ vs $a$ plot is given in Fig.~\ref{psiplot}. One can clearly infer that $\Psi$ evolves into negative negative values from a positive domain, at late times. We note, speculatively, that the universe has a proclivity to move towards thermodynamic equilibrium as the universe keeps on expanding. This is also expected as far as a standard $\Lambda$CDM cosmology is concerned and marks a point of cosmological consistency. 
As a summary, this section introduces a very simple formalism/scheme to compose an \textit{intended} cosmological behavior supported by observations. In a sense, it avoids the complication of taking a non-linear problem head-on, i.e., it avoids the compulsion of working out an exact solution. The dynamics fulfils primary necessities, for instance, consistent growth pattern of matter over-density, compatible evolution patterns of the deceleration profile of the universe and also suggests a thermodynamic equilibrium. No reference to a modified gravity theory is needed as this exercise is simply based on a credible set of observational data and their analysis. This gives us more freedom to solve the modified field equations. We can now directly solve for the field profiles and comment on the desired structure of an interacting scalar-fermion theory that can support this cosmological system. We discuss this in the next section.

\section{A Reconstruction of the Interacting Boson-Fermion Model}
As discussed in Section $2$, the spatially homogeneous cosmology in an interacting scalar-fermion setup is given by the field Eqs.~(\ref{kg1}), (\ref{kg2}) and (\ref{kg3}). This section ellaborates the numerical solutions to this set of equations, using the closed form of Hubble as in Eq.~(\ref{hubbleansatz}). The rest of the mathematical manipulation is quite simple, we change the argument of these equations from cosmic time $t$ and write all the components as functions of the redshift.

\begin{figure}[t]
\begin{center}
\includegraphics[angle=0, width=0.40\textwidth]{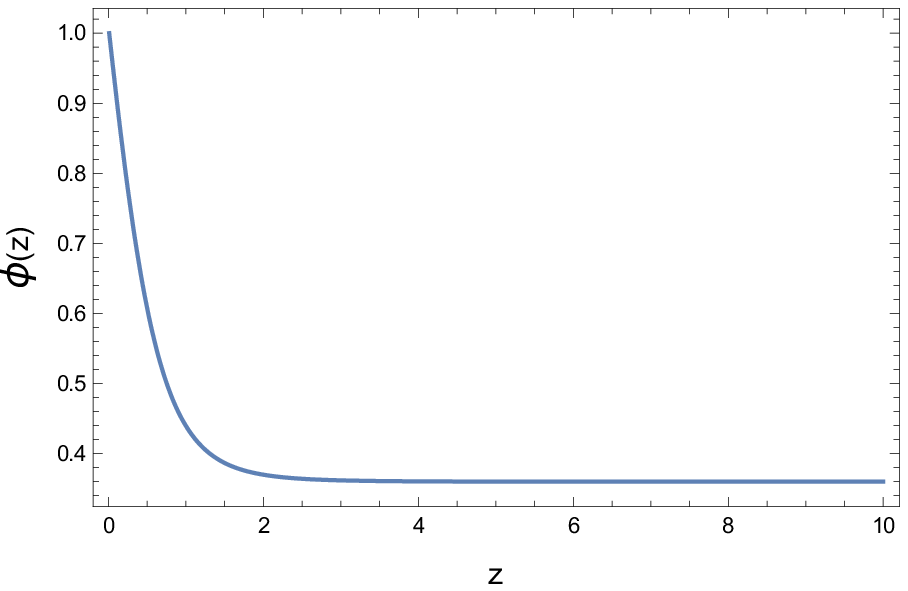}
\includegraphics[angle=0, width=0.40\textwidth]{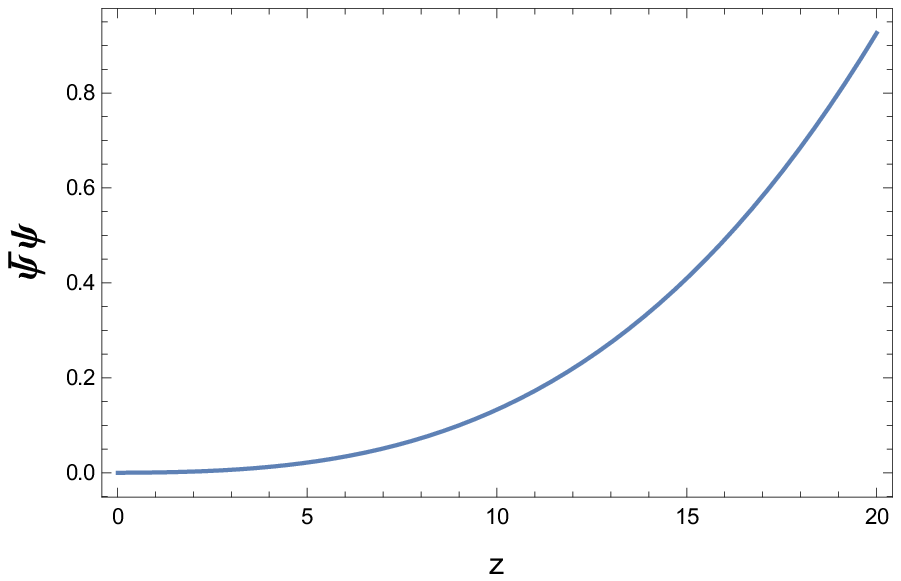}
\caption{Profile of the scalar field $\phi$ and the bilinear $\bar{\psi}\psi$ as functions of redshift $z$, for the best-fit parameter values of Hubble, $\lambda_0$ and $\delta$. The boson self-interaction is assumed to be Higgs-type for this Figure.}
\label{fields}
\end{center}
\end{figure}

\begin{figure}
\begin{center}
\includegraphics[angle=0, width=0.40\textwidth]{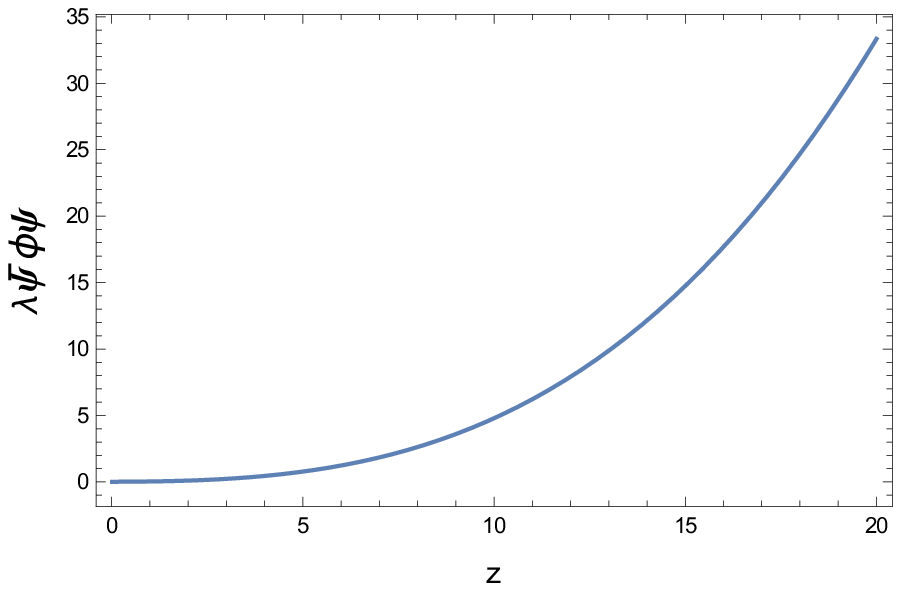}
\includegraphics[angle=0, width=0.40\textwidth]{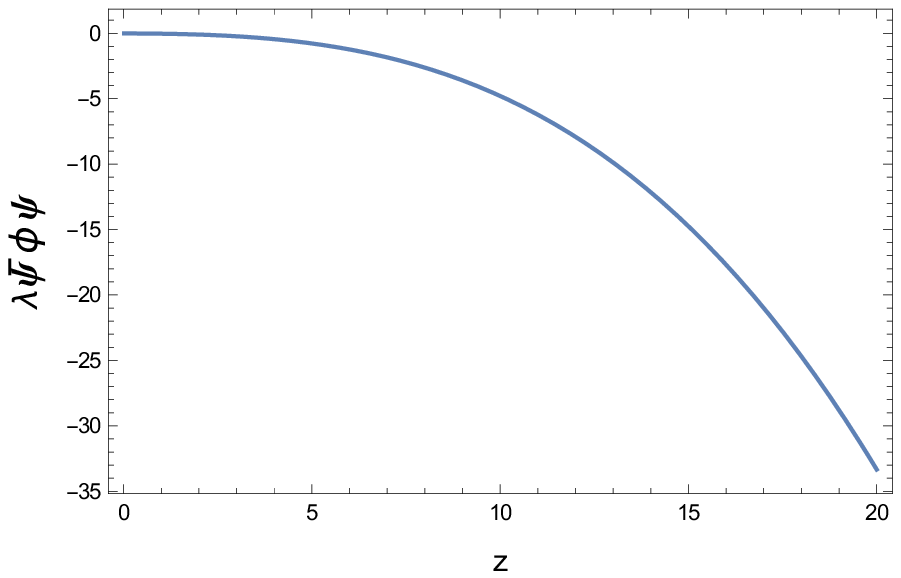}
\caption{Profile of the Yukawa interaction as a function of redshift, for the best-fit parameter values of Hubble, $\lambda_0$ and $\delta$. The boson self-interaction is assumed to be Higgs-type for this Figure. The graph on the top panel is for $\lambda > 0$ and on the bottom panel is for $\lambda < 0$.}
\label{yukawa}
\end{center}
\end{figure}

\begin{align}\label{1eq}
H(z) &= H_{0}\left[1 + \lambda_{0} (1+z)^{\delta} \left\lbrace (1+z)^{3} - 1 \right\rbrace\right]^{\frac{1}{2}}, \\\notag
\dot{\phi} &= \frac{d\phi}{dz} \frac{dz}{da}\frac{da}{dt}, \\\notag
&= -H_{0}\phi'(1+z)\left[1 + \lambda_{0} (1+z)^{\delta} \left\lbrace (1+z)^{3} - 1 \right\rbrace\right]^{\frac{1}{2}}, \\&&
\end{align}
where we have written $d\phi/dz$ as $\phi'$ and the scale factor as $1/(1+z)$. The second derivative of $\phi$ can be written as
\begin{align}\notag
\ddot{\phi} &= H_{0}^{2}(1+z)^{2} \phi''\Big[ 1 + \lambda_{0}(1+z)^{\delta}\Big\lbrace (1+z)^3 - 1\Big\rbrace \Big] \\\notag
&\quad + H_{0}^{2}(1+z)\phi' \Big[ 1 + \lambda_{0}(1+z)^{\delta}\Big\lbrace (1+z)^3 - 1\Big\rbrace \Big] \\\notag
&\quad +\frac{H_{0}^{2}}{2}(1+z)^{2}\phi' \Big[\delta \lambda_{0}(1+z)^{\delta - 1}\Big\lbrace (1+z)^{3} - 1\Big\rbrace \\
\label{2eq}
&\quad + 3\lambda_{0}(1+z)^{\delta+2}\Big]\,.
\end{align}
Also, note that the exercise in this section is only for the best fit parameter values of $H_{0}$, $\lambda_0$ and $\delta$. We use Eqs. (\ref{1eq}) and (\ref{2eq}) in the three field equations and solve for the unknowns, namely, the scalar field $\phi$, the fluid energy density $\rho_{m}$, and the pressure $p_{m}$. To begin with, we take the scalar self-interaction to be of Higgs-type,
\begin{equation}\label{eq:VHiggs}
V(\phi) = \frac{\mu^{2}}{2}\phi^{2} + \frac{\lambda_h}{4}\phi^{4}.
\end{equation}
$\mu^2$ is negative and has a dimension of mass squared, while $\lambda_h$ is dimensionless. 
We note that even though $\phi$ is taken to have a Higgs-like self-interaction, it is not the Higgs field, so its parameters are not those of the Standard Model. 
Fig.~\ref{fields} shows the evolution of the scalar field $\phi$ (top panel) and the bilinear $\bar{\psi}\psi$ (bottom panel). It is apparent that the fermion density (bilinear) becomes subdued during a late-time acceleration and the scalar starts to dominate. Fermion density becomes almost negligible in the present era. The reverse of this is true for an epoch of deceleration ($z > 1$) and therefore a cosmological deceleration is more of a fermion-driven phenomenon with bosonic components staying subdued. To drive this switchover of roles between the fields one requires a mediator. The Yukawa interaction term can be imagined in such a role, which goes down alongwith the fermion density, i.e., dominates the deceleration but decays with a cosmic acceleration. We recall that $\mathcal{L}_{\rm Yukawa}(\phi,\psi) = -\lambda\overline\psi\phi\psi$ is defined as the Yukawa interaction and show the profile of the interaction in Fig. \ref{yukawa} for two signatures of the Yukawa parameter, namely, positive (top panel) and negative (bottom panel). \\

\begin{figure}[t]
\begin{center}
\includegraphics[angle=0, width=0.40\textwidth]{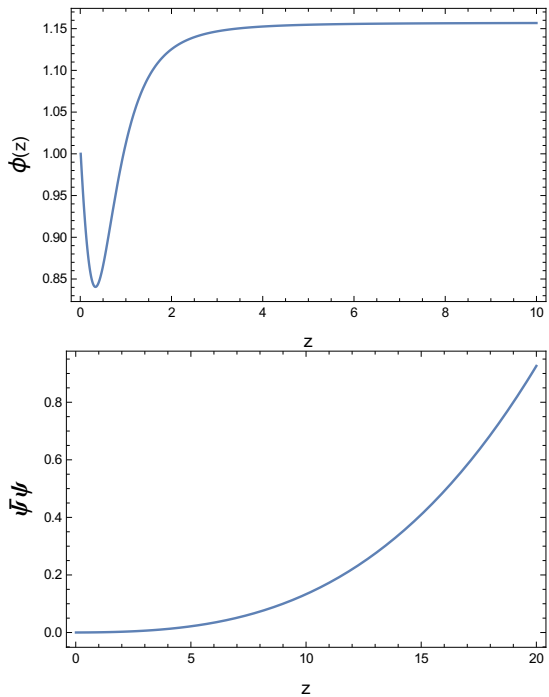}
\includegraphics[angle=0, width=0.40\textwidth]{bilinear.eps}
\caption{Profile of the scalar field $\phi$ and the bilinear $\bar{\psi}\psi$ as functions of redshift $z$, for the best-fit parameter values of Hubble, $\lambda_0$ and $\delta$. The bosonic self-interaction is assumed to be an inverse-power law type for this Figure.}
\label{fields1}
\end{center}
\end{figure}

Qualitatively, this profile is novel, as a change in the scalar self-interaction affects the pattern negligibly. This can be checked by using different cosmologically relevant forms of the potential, such as, an exponential or a simple power law form. The only notable difference comes when we choose to work with an inverse power law scalar self-interaction with a runaway form,
\begin{equation}\label{inv}
V(\phi) = M^{4+n}\phi^{-n},
\end{equation}
where $n > 0$ and $M$ is of mass dimension. These potentials were first popularized in a cosmological context as \textit{tracking quintessence} fields by Ratra and Peebles \cite{ratra} and also by Steinhardt et al.~\cite{steinhardt}. The potential plays a particularly crucial role in the so-called chameleon scalar field theories and the subsequent resolution of the Equivalence Principle violation problems~\cite{khoury}. In Fig. \ref{fields1}, we plot the numerical solutions of the field Eqs. (\ref{kg1}), (\ref{kg2}) and (\ref{kg3}), for the new choice of potential as in Eq. (\ref{inv}). In this case, while the late-time evolution of the universe remains scalar-dominated, the profile of the scalar shows a curious formation of minima just before the onset of late-time acceleration, i.e., around the transition redshift $z \sim 1$. For higher redshifts, the scalar varies negligibly and behaves effectively as a constant. The fermion density as well as the Yukawa interaction behaves in the same manner as compared to the example of Higgs scalar interaction, apart from an overall scaling in the profile.  \\

\begin{figure}
\begin{center}
\includegraphics[angle=0, width=0.40\textwidth]{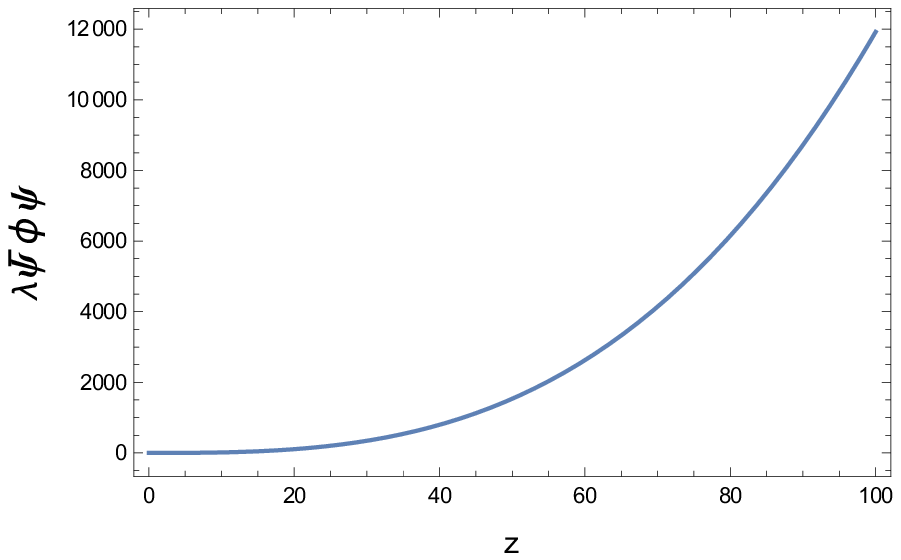}
\includegraphics[angle=0, width=0.40\textwidth]{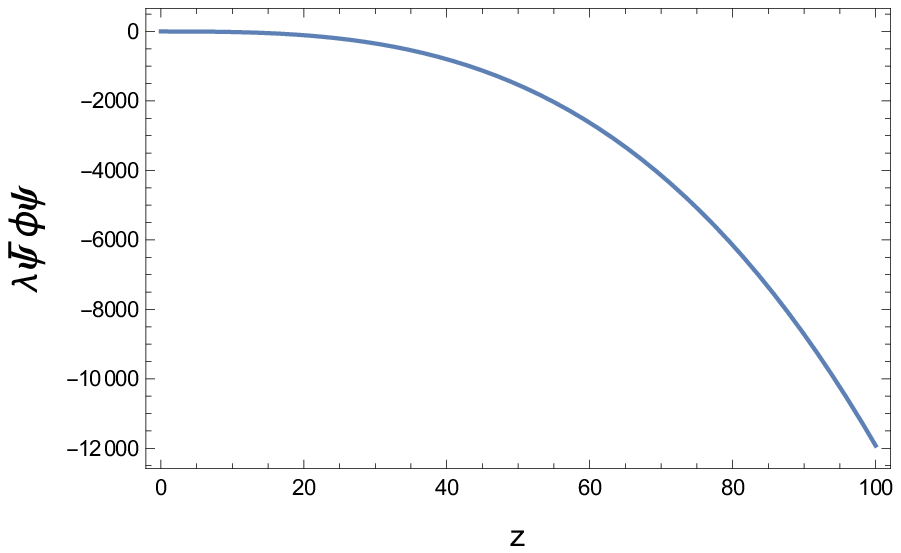}
\caption{Profile of the Yukawa interaction potential as a function of redshift, for the best-fit parameter values of Hubble, $\lambda_0$ and $\delta$. The bosonic self-interaction is assumed to be an inverse-power law type for this Figure. The graph on the top panel is for $\lambda > 0$ and on the bottom panel is for $\lambda < 0$.}
\label{yukawa1}
\end{center}
\end{figure}

\begin{figure}
\begin{center}
\includegraphics[angle=0, width=0.40\textwidth]{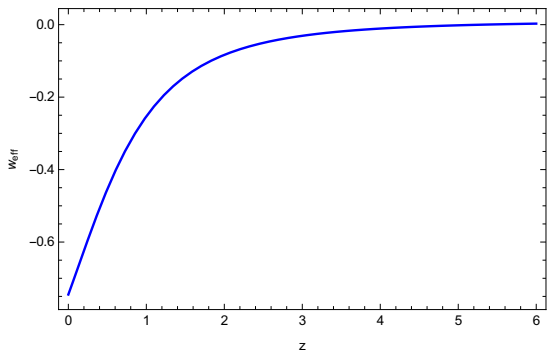}
\caption{Profile of the Effective Equation of State of the cosmological system, for the best-fit parameter values of Hubble, $\lambda_0$ and $\delta$.}
\label{EOS}
\end{center}
\end{figure}

We conclude this section by showing how the effective EOS of this cosmological system of interacting boson-fermions can evolve as a function of redshift. It is relatively straightforward to write $\rho_{eff} \sim \rho_{m} + \rho_{mod} \sim 3H^2$ and $p_{eff} \sim -2\dot{H} - 3H^{2}$ using the field equations. Note that all the modified terms are contained within the effective fluid density and pressure, while the rest of the terms are just simply functions of Hubble. Ratio of these two can be written as $w_{eff}$, and is plotted in Fig. \ref{EOS} as a function of redshift. We note that the system behaves as a Dark Energy dominated system during the present epoch ($w_{eff} \sim -1$). During earlier epochs, i.e., for $z > 1$, $w_{eff}$ approaches a zero value, indicating a dust/cold dark matter dominated deceleration. 

\section{A possibility of Unified Cosmic Expansion History}
\begin{figure}
\begin{center}
\includegraphics[width=0.40\textwidth]{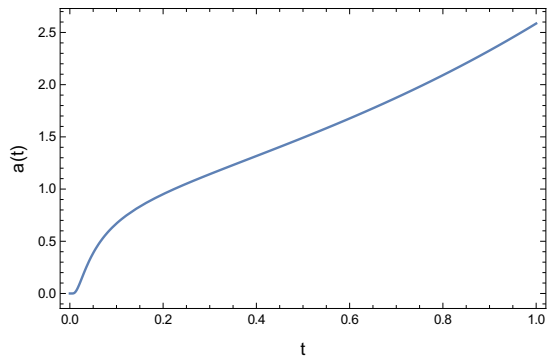}
\caption{Profile of the envisaged Scale factor as a function of cosmic time. The model parameters are chosen as $H_{0} = 1, H_{1} = 0.05, n = 4$ and $a_{0} = 1$.}
\label{unifiedplot}
\end{center}
\end{figure}

It is entirely possible that this system driven by interacting boson and fermion can be a viable driver of the early universe cosmology as well. As a natural intuition, we think about a unified cosmic time history only at the level of a simple toy model. Indeed, most of the existing cosmological models depict simple patches of cosmic time and fit them with some era-specific observational data-sets. The complete time history of the universe, in a broad sense, should simply start with an early inflation right after the Big Bang. It should then be followed by an extended phase of deceleration to allow the large scale structures to develop. Finally it should enter the late-time acceleration. This involves more than one instance of transition of the universe from deceleration into acceleration and vice-versa and the cosmological scale factor should behave accordingly. 
We therefore consider the following ansatz, 
\begin{align}\label{H.evolve}
H(t) &= H_{0}+\frac{H_{1}}{t^{n}} \\
a(t) &= a_{0}\exp \Big[H_{0}t -\frac{H_{1}}{(n-1)t^{n-1}}\Big]\,.
\label{a.evolve}
\end{align}
We have introduced the parameters $H_{0}$, $H_{1}$, $n$ and $a_{0}$ at the outset. $t \sim 0$ is the point where the universe begins to expand and thereafter $t$ always remains positive. In Fig.~\ref{unifiedplot}, we show how this scale factor evolves with cosmic time, for the choice of parameters
\begin{equation}
H_{0} = 1, \,\ H_{1} = 0.05, \,\ n = 4, \,\ a_{0} = 1\,.
\end{equation}
As far as the parameters are positive, any value of them can generate the desired dynamics. However, we have chosen this particular set to clearly illustrate different phases within a finite scale. For these values, $0 \lesssim t \lesssim 0.05$ is the era of an early acceleration prior to the subsequent epochs of deceleration and late-time acceleration. \\

A cosmic transition can be realized from the effective equation of state (EOS) defined as
\begin{align}
w_{\rm eff} &= -1-\frac{2\dot{H}}{{3H}^{2}}\,, \notag\\
&=-1+\frac{2nH_{1}t^{n-1}}{\Bigg(H_{0}t^{n}+H_{1}\Bigg) ^{2}}\,.
\label{1.14}
\end{align}
We note an interesting point. There are two epochs of acceleration, the first for $t \sim 0$ (early inflation) and another in the large $t$ limit (late-time acceleration). In both of the cases, we find that $w_{\rm eff} \rightarrow -1$, which signifies an effective negative pressure. In other words, there exists a few \textit{`critical points'} where the universe makes a smooth transition from one epoch into the next. These transitions can be understood from the deceleration parameter $q = -\frac{\ddot{a}a}{\dot{a}^2}$. The sign of this parameter depends on the sign of the acceleration $\frac{\ddot a}{a}$, written as
\begin{equation}
\frac{\ddot a}{a} = \dot{H} +H^{2} = -\frac{nH_{1}}{t^{n+1}} + \left( H_{0}+\frac{H_{1}}{t^{n}}\right)^{2}. \label{1.15}
\end{equation}
A transition is exactly given by a zero value of $\ddot{a}/a$. We calculate the time of transition in terms of the model parameters as
\begin{equation}
t_{\pm} \approx \left[ \sqrt{nH_{1}} \,\, \frac{\left( 1\pm
\sqrt{1-\frac{4H_{0} }{n}} \, \right)}{2H_{0}} \, \right]^{2/n}.
\label{1.16}
\end{equation}
One must ensure that $\frac{4H_{0}}{n} \leq 1$ to avoid imaginary values of time. We have chosen $H_{0} = 1$ and $n = 4$, which leads to $\sqrt{1-\frac{4H_{0} }{n}} = 0$. For a different set of model parameters, the critical point formations will be realized at a different set of values of cosmic time. Therefore, broadly the time history of the universe becomes,
\begin{itemize}
\item {Early acceleration $\sim 0 < t < t_{-}$.}
\item {Deceleration $\sim t_{-} < t < t_{+}$.}
\item {Late-time acceleration $\sim t > t_{+}$.}
\end{itemize}
\begin{figure}
\begin{center}
\includegraphics[width=0.40\textwidth]{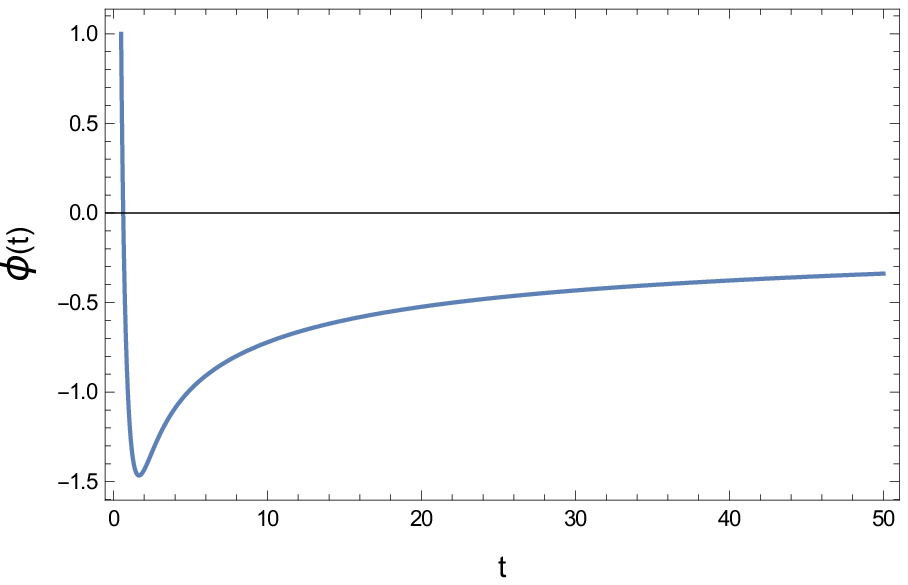}
\includegraphics[width=0.40\textwidth]{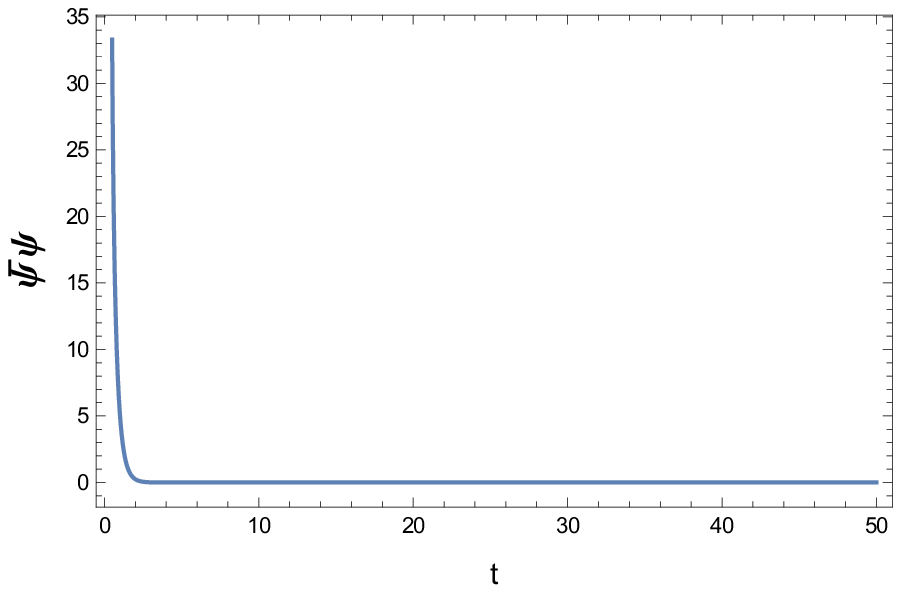}
\caption{Profile of the boson scalar and the bilinear for a unified time history of the universe as a function of cosmic time. The model parameters are chosen as $H_{0} = 1, H_{1} = 0.05, n = 4$ and $a_{0} = 1$. A Higgs-like scalar self-interaction potential is chosen for this Figure.}
\label{unifiedplot2}
\end{center}
\end{figure}

This is not yet a complete model. It simply establishes the fact that the chosen solution gives a particular cosmic behavior. Observations need to constrain different parameters. Leaving observational constraints aside, choice of the parameters should constrain the matter distribution in a given model and we try to discuss this next. We now solve the field equations of the interacting scalar-fermion theory, i.e., Eqs.~(\ref{kg1}), (\ref{kg2}) and (\ref{kg3}) using Eq.~(\ref{H.evolve}). The numerical solutions give us the profiles of the boson, the fermion density and the Yukawa interaction as a function of cosmic time. The evolution remains qualitatively the same if we change the values of the model parameters other than an overall scaling. However, it shows some interesting difference for different choices of the scalar self-interaction. First, we solve the system for a Higgs-like scalar self-interaction and plot the evolution of the scalar field and the fermion density in Fig.~\ref{unifiedplot2}. The fermion density is plotted for a positive value of the parameter $C \sim 100$, which is directly related to the mass scale of the fermion. In our present scale, the time history of the universe is realized within a range of cosmic time $0 \lesssim t \lesssim 2$ in units $H^{-1}$. This leads to a general deduction that in the early universe, both of the fields dominate the cosmology. An onset of the subsequent epoch of deceleration sees both of the fields starting to decay, however, with different rates. While the fermion density falls of drastically and becomes almost negligible during late times, the scalar field asymptotically reaches a constant value. There is a time between the early acceleration and the subsequent deceleration, when the scalar field changes its sign from positive to negative. This is followed by a formation of minima of the scalar field. \\

\begin{figure}
\begin{center}
\includegraphics[width=0.40\textwidth]{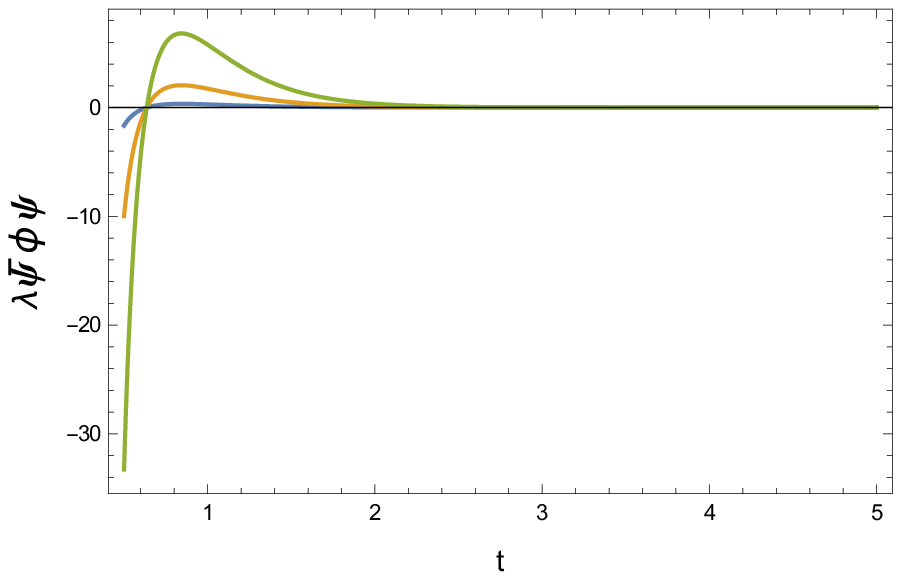}
\includegraphics[width=0.40\textwidth]{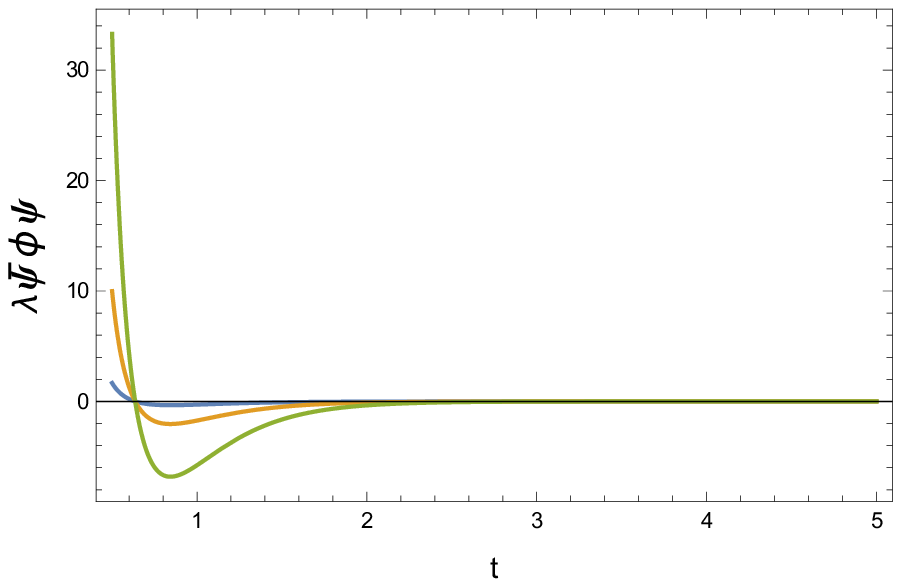}
\caption{Profile of the Yukawa interaction for a unified time history of the universe as a function of cosmic time. The model parameters are chosen as $H_{0} = 1, H_{1} = 0.05, n = 4$ and $a_{0} = 1$. A Higgs-like scalar self-interaction potential is chosen for this Figure. (i) Top Panel : The Yukawa parameter $\lambda < 0$. (ii) Bottom Panel : The Yukawa parameter $\lambda > 0$.}
\label{unifiedplot3}
\end{center}
\end{figure}

This evolution has a clear contribution from the Yukawa interaction between the fields, as shown in Fig.~\ref{unifiedplot3}. The three different plots of the Yukawa correspond to three different values of the Yukawa parameter $\lambda$. We remember that the Yukawa interaction comes in as $\mathcal{L}_{\rm Yukawa}(\phi,\psi) = -\lambda\overline\psi\phi\psi$ and the parameter $\lambda$ contributes into the evolution as $\lambda C$, with the fermion density being given by $\overline\psi\psi=\frac{C}{a^3}$. The top plot of Fig. \ref{unifiedplot3} is for three values of $\lambda C$, chosen as $-5, -30, -100$, i.e, $\lambda$ is chosen to be negative. The bottom plot of Fig. \ref{unifiedplot3} is for three values of $\lambda C$, chosen as $5, 30, 100$, i.e, $\lambda$ is chosen to be positive. For the negative values of $\lambda C$, the Yukawa interaction is negative and dominating during the early acceleration. It moves into a positive domain as the epoch of deceleration sets in. The larger the value of $\lambda C$ is, more extended the epoch of deceleration becomes. Since the Yukawa interaction is linearly related to the scalar field, it crosses zero from a positive into the negative domain exactly when the scalar field crosses. We can comment that the choice of the parameter $\lambda C$ dictates the strength of the Yukawa interaction during an early universe. For the positive values of $\lambda C$, the interaction is exactly the opposite. As the universe begins to move into a late-time acceleratton, the Yukawa interaction becomes increasingly negligible.  \\

\begin{figure}
\begin{center}
\includegraphics[width=0.40\textwidth]{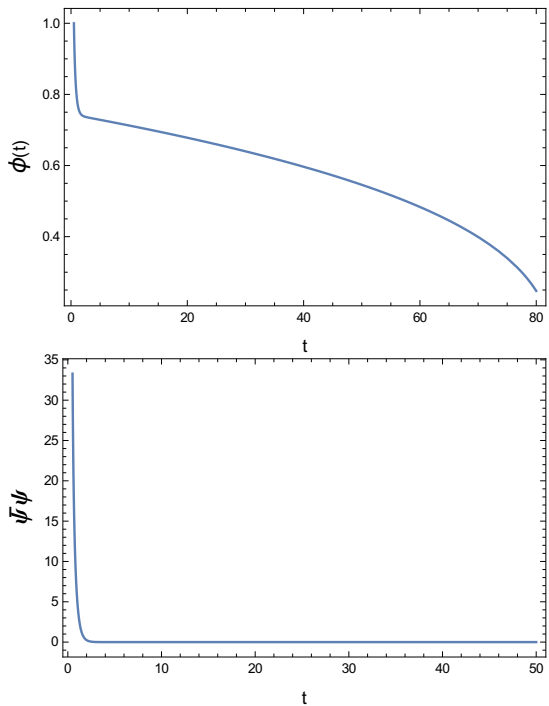}
\includegraphics[width=0.40\textwidth]{bilinear_unified.eps}
\caption{Profile of the boson scalar and the bilinear for a unified time history of the universe as a function of cosmic time. The model parameters are chosen as $H_{0} = 1, H_{1} = 0.05, n = 4$ and $a_{0} = 1$. An inverse power-law scalar self-interaction potential is chosen for this Figure.}
\label{unifiedplot4}
\end{center}
\end{figure}

\begin{figure}
\begin{center}
\includegraphics[width=0.40\textwidth]{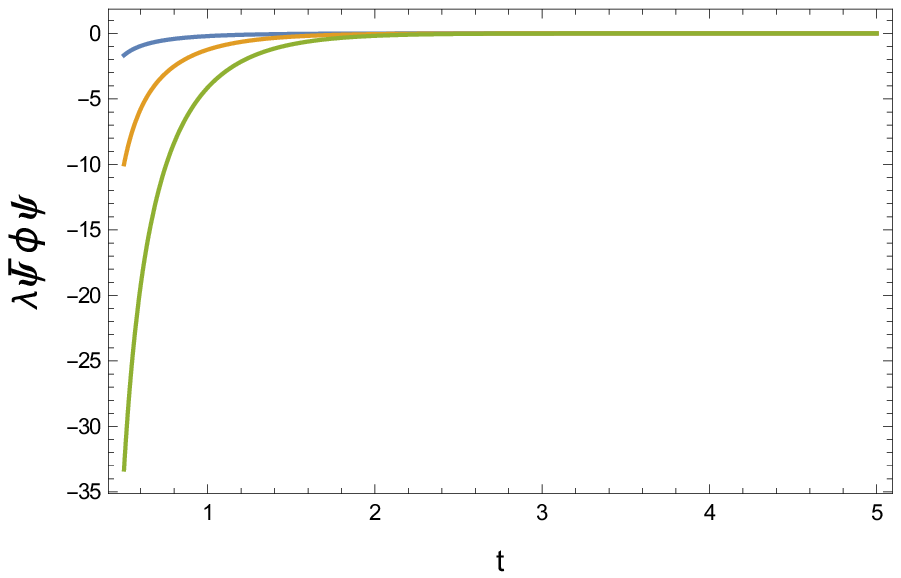}
\includegraphics[width=0.40\textwidth]{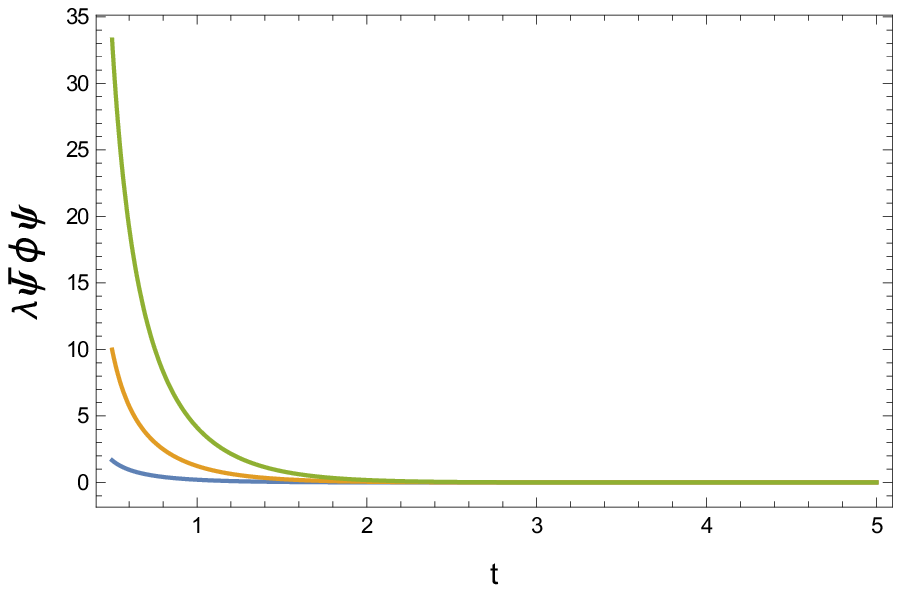}
\caption{Profile of the Yukawa interaction for a unified time history of the universe as a function of cosmic time. The model parameters are chosen as $H_{0} = 1, H_{1} = 0.05, n = 4$ and $a_{0} = 1$. An inverse power-law scalar self-interaction potential is chosen for this Figure. (i) Top Panel : The Yukawa parameter $\lambda < 0$. (ii) Bottom Panel : The Yukawa parameter $\lambda > 0$.}
\label{unifiedplot5}
\end{center}
\end{figure}

We repeat the same exercise once more, only this time making the scalar self-interaction to be an inverse power-law type, as in Eq. (\ref{inv}). The numerical solution for the scalar is shown as a plot in Fig. \ref{unifiedplot4} (top panel). It shows that although the scalar field always remains positive throughout the cosmic evolution, it always decays with cosmic time. The decay profile is quite rapid during the early inflation. Eventually, it slows down and the scalar becomes a monotonically decreasing function. The bottom panel shows the evolution of fermion density, which is exactly similar as its profile was in a Higgs scalar potential example. Fig. \ref{unifiedplot5} shows the plot of the Yukawa potential. Once again, the three different plots correspond to three different values of the parameter $\lambda C$. The top plot of Fig. \ref{unifiedplot5} is for three negative values of $\lambda C$, chosen as $-5, -30, -100$. The bottom plot of Fig. \ref{unifiedplot5} is for three positive values of $\lambda C$, chosen as $5, 30, 100$. For the negative values of $\lambda C$, the Yukawa interaction is negative throughout the cosmic expansion. There is no instance of crossing into a positive domain as in the previous example of Higgs scalar interaction. The interaction dominates during the early acceleration and starts decaying rapidly, as an era of deceleration sets in. The larger the value of $\lambda C$ is, more extended the epoch of deceleration becomes. For the positive values of $\lambda C$, the evolution of the interaction behaves, once again, in an exact opposite manner. As the universe accelerates into the future, the Yukawa interaction becomes increasingly negligible irrespective of the choice of this parameter.  

\section{Conclusion}
In this manuscript we have described a simple way to adjudge the viability of a non-standard cosmological construct, involving an interacting assembly of bosonic and fermionic fields, based on a set of credible astrophysical observations. The fields are self-interacting and they also interact amongst themselves through a Yukawa interaction term. In order to accommodate the accelerated expansion of the universe this setup must exhibit a few crucial constraints on the field profiles. We have discussed these constraints using cosmological solutions with evolving Equation of State, such that the universe can always allow a smooth transition between different epochs. The scalar-fermion Yukawa interaction plays a particularly interesting role in this scenario. It drives the fermion density with cosmic time and can be tagged as as an effective mediator between different epochs of cosmic expansion. \\

We infer a desirable cosmological behavior of the universe from standard cosmological data-sets such as the Joint Light Curve Analysis of the Supernovae data, Hubble parameter measurements and the Baryon Accoustic Oscillation Data. This analysis is a straightforward method to reverse engineer the best possible structure of a theory from cosmological requirements. We parametrize the $Om(z)$ parameter, the present matter density contrast of the universe in the process and derive an observationally viable equation for the Hubble function. We numerically solve the field equations in order to reconstruct the scalar-fermion theory. This reconstruction serves two purposes. First, a straightforward parametrization of $Om(z)$ provides fresh ideas regarding departures in the estimation of matter density constrast. Secondly and more importantly, this scheme has observational viability due to astrophysical data analysis and is independent of the particular theory of gravity chosen at the outset.   \\

As a general and expected conclusion we find that the late-time acceleration of the universe is dominated by a sef-interacting scalar field. Any fermion density present during the preceding epoch of deceleration falls off sharply by the time the acceleration sets off. The Yukawa interaction, being directly proportional to the Fermion density, follows suit and starts decaying around the redshift of transition, i.e., $z_{t} \sim 1$. We derive the structures for a Higgs-like and an inverse power law self-interaction potential of the scalar field. However, the bilinear as well as the Yukawa interaction profiles remain unaffected by any change in the scalar self-interaction. Any possible modification to the theoretical structure is contained within the profile of the scalar field itself (see Fig. \ref{fields} and Fig. \ref{fields1} for reference). We also discuss the compatibility of the theory in a unified cosmic expansion time history and in particular, in an early universe cosmology. We work with an ansatz, rather intuitively, that follows an itinerary of \textit{early inflation $\rightarrow$ extended deceleration $\rightarrow$ late-time acceleration}. It successfully generates a varying EOS and can exhibit more than one smooth transitions between different cosmological epochs of acceleration/deceleration. Existence of definite points of phase transitions of the universe, called the \textit{critical time} estimates are discussed. To support this mathematical evolution the theory should have some specific constraints on its constituent fields and their interactions, which we discuss at some length. We show that the Fermion density dominates over the boson during an early acceleration. Throughout the epoch of an extended deceleration it gradually falls off towards a negligible scale and mildly varying nature. The scalar field dominates the subsequent epoch of late-time acceleration and the manner of dominance, i.e., the present strength, dominance of the time derivatives are decided by its self-interaction (Higgs-type or inverse power-law). The Yukawa interaction governs the mutual evolution of these fields in a very interesting manner. Immediately after the Big Bang, the interaction is at an extrema. If the Yukawa parameter $\lambda > 0$, then this extrema is a minima with a negative value and if $\lambda > 0$, it is a minima with a negative value. If $\lambda < 0$, the extrema is a maxima with a positive value. The interaction starts loosing strength as the Universe evolves towards deceleration. The onset of deceleration is signified by the formation of a second extrema of the interaction, which is either a minima in the negative domain or a maxima in the positive domain, depending on $\lambda$. Subsequently, the strength of the interaction starts decaying steadily and asymptotically reaches a negligible profile during the present acceleration. For an inverse power self-interaction of the scalar field, the second extrema of the Yukawa profile is not clearly understood, however, qualitatively the interaction profile behavior behaves in a similar fashion. \\

Theoretical science often involves a great deal of story-telling, based on credible mathematics and experiments. It is true that the field of theoretical cosmology is almost saturated with different candidates that are the \textit{usual suspects} behind a cosmic acceleration. A fermion driven cosmology is just a different story full of new possibilities. It provides, in the present form, the simplest possible toy model of the universe and its constituent fields and fluids. More importantly, it leaves open a few doors for the readers to consider extended versions of an interacting scalar-fermion theory. A few of such possibilities include interacting fermions in scalar-tensor theories, interacting fermions in a theory with non-zero torsion, and hopefully these will be discussed by the authors in the near future. 

\section*{Data Availability}
The authors confirm that all relevant data, included in the manuscript, are either from public domain or from published papers, all of which are duly cited. 

\section*{References}


\end{document}